\documentclass[twocolumn,showpacs,preprintnumbers,amsmath,amssymb,superscriptaddress]{revtex4}
\usepackage{placeins}
\usepackage{graphicx}
\usepackage{amsmath}
\usepackage{times}
\usepackage{color}

\begin{document}

\draft

\title{Hypergraph model of social tagging networks}

\author{Zi-Ke Zhang}
\thanks{zhangzike@gmail.com}
\affiliation{Department of Physics, University of Fribourg, Fribourg
CH-1700, Switzerland}

\author{Chuang Liu}
\thanks{liuchuang@mail.ecust.edu.cn}
\affiliation{Department of Physics, University of Fribourg, Fribourg
CH-1700, Switzerland}
\affiliation{School of Business, East China
University of Science and Technology, Shanghai 200237, P. R. China}
\affiliation{Engineering Research Center of Process Systems
Engineering (Ministry of Education), East China University of
Science and Technology, Shanghai 200237, P. R. China}

\begin{abstract} \textnormal{\small {The past few years have witnessed
the great success of a new family of paradigms, so-called
folksonomy, which allows users to freely associate tags to resources
and efficiently manage them. In order to uncover the underlying
structures and user behaviors in folksonomy, in this paper, we
propose an evolutionary hypergrah model to explain the emerging
statistical properties. The present model introduces a novel
mechanism that one can not only assign tags to resources, but also
retrieve resources via collaborative tags. We then compare the model
with a real-world dataset: \emph{Del.icio.us}. Indeed, the present
model shows considerable agreement with the empirical data in
following aspects: power-law hyperdegree distributions, negtive
correlation between clustering coefficients and hyperdegrees, and
small average distances. Furthermore, the model indicates that most
tagging behaviors are motivated by labeling tags to resources, and
tags play a significant role in effectively retrieving interesting
resources and making acquaintance with congenial friends. The
proposed model may shed some light on the in-depth understanding of
the structure and function of folksonomy.}}
\end{abstract}


\maketitle

\section{Introduction}
Networks provide us a powerful and versatile tool to recognize and
analyze complex systems where nodes represent individuals, and links
denote the relations between them. Recently, many efforts have been
addressed in understanding the structure, evolution and dynamics of
complex networks \cite{Albert2002, Dorogovtsev2002, Newman2003,
Boccaletti2006, Costa2007}. The advent of Web 2.0 and its affiliated
applications bring a new form of user-centric paradigm which can not
be fully described by pre-existing models on unipartite or bipartite
networks. One such example is the user-driven emerging phenomenon,
\emph{folksonomy}, which allows users to upload resources
(bookmarks, photos, movies, etc.) and freely assign them with
user-defined words, so-called \emph{tags}. Folksonomy requires no
specific skills for user to participate, broadens the semantic
relations among users and resources, and eventually achieves its
immediate success in a few years. Presently, a large number of such
applications can be found online, such as \emph{Del.icio.us}
\cite{f1}, \emph{Flickr} \cite{f2}, \emph{CiteULike} \cite{f3}, etc.
With the help of those platforms, users can not only store their own
resources and manage them with collaborative tags, but also look
into other users' collections to find what they might be interested
in by simply keeping track of the baskets with tags. Unlike
traditional information management methods where words (or indices)
are normally pre-defined by experts or administrators, e.g. the
library classification systems, a tagging system allows users to
create arbitrary tags that even do not exist in dictionaries.
Therefore, those user-defined tags can reflet user behaviors and
preferences with which users can easily make acquaintance,
collaborate and eventually form communities with others who have
similar interests \cite{Sen2006}.

Up to now, a variety of research works have been done in realizing
the structure and dynamic process of folksonomy. Golder \emph{et
al.} studied the usage patterns of \emph{collaborative tagging
systems} and classified seven kinds of tag functions
\cite{Golder2006}, which is very helpful for us in better
understanding both the user behaviors and tagging purposes. In
addition, the keywords or PACS numbers based methods are put forward
to reveal the underlying structure of co-authorship and citation
networks \cite{Palla2008, ZhangZK2008}. Furthermore, many efforts
have been done to explain how folksonomy emerges. Cattuto \emph{et
al.} \cite{Cattuto20071} investigated the dynamics of an open-ended
system with a memory-based Yule-Simon model. The model considered
the aging effect of tags, as well as the frequency of tag
occurrence. In Ref. \cite{Lambiotte2006}, they tried to model
folksonomy in a form of tripartite graphs.

Recently, the \emph{hypergraph} theory \cite{Karypis1997} allows a
hyperedge to connect an arbitrary number of vertices instead of two
in regular graphs. Therefore, it provides us a promising way to
better understand a wide range of real systems. Up to now, it has
been found applications in \emph{Personalized Recommendation}
\cite{Blattner2009, ZhangZK2010, ShangMS2010}, \emph{ Population
Stratification} \cite{Vazquez2008}, and \emph{Cellular Networks}
\cite{Klamt2009}, etc. Besides, the definition is comparatively
appropriate to uncover underlying usage patterns and essential
structures of folksonomies. Ghoshal \emph{et al.} \cite{Ghoshal2009}
proposed a random hypergraph model to represent the ternary
relationship where a hyperedge consists of one user, one resource
and one tag, and reproduced many properties of folksonomy by the
model. Zlati\'{c} \emph{et al.} \cite{Zlatic2009} extensively
defined a number of useful topological features based on hypergraph
representation, which can be considered as a standard tool in
understanding the structure of tagged networks.

\begin{figure}[ht]
\center\scalebox{0.8}[0.8]{\includegraphics{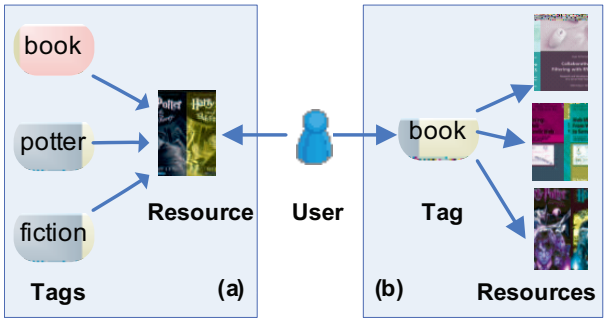}}
\caption{(Color online) Illustration of two typical user tagging
behaviors: (a) the user finds a resource (e.g. a book) via web
surfing and annotate it with three tags for further use; (b) s/he
collects one or some books by filtering out unrelated information
with the tag 'book'. }\label{model}
\end{figure}

In this paper, we propose a hypergraph model to illustrate the
emergence of some statistical properties in folksonomy, including
degree distribution, clustering coefficients and average distance
between nodes. We consider two typical user tagging behaviors: (i)
one might be aware of a resource via web surfing or word-of-mouth
propagation, and then save it as his/her own favorite collection and
annotate it with several tags of related topics for efficient
management and retrieval; (ii) s/he might firstly pick up one or
several compound tags, and then choose one possible resource from
the retrieval results. Recently, a considerable amount of researches
have focused on the previous motivation \cite{Halpin2007,
Cattuto2009}, while the latter one is comparatively lack of
attention. Actually, tag is able to provide more relevant results
according to its simple yet essential property of collaboration and
semantics. Fig. \ref{model} shows those two different kinds of
mechanisms.

In this model, users can manage resources with collaborative tags,
and find resources by tags via serendipitous browsing. We compare
the model to one real-world dataset, \emph{Del.icio.us}, and find
good agreement between them.

\section{Modeling Tripartite Hypergraphs}

\begin{figure}[ht]
\center\scalebox{0.30}[0.3]{\includegraphics{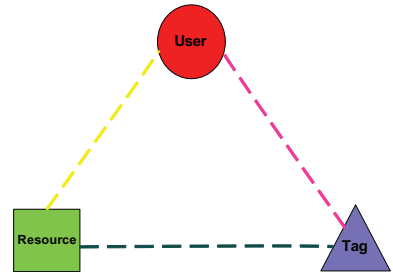}}
\caption{(Color online) A hyperedge illustration of the basic unit
in our network. There are three types of vertices in each hyperedge
(represented as a triangle), depicted by one red circle, one green
rectangle and one blue triangle which respectively represent a user,
a resource and a tag in folksonomy.}\label{basic}
\end{figure}

We begin our study with some related definitions of tripartite
hypergraph that we will analyze. In this paper, we use the
tripartite hypergraph representation given by Ref.
\cite{Ghoshal2009}, where a hyperedge simply consists of one user,
one resource and one tag. Fig. \ref{basic} gives a visual
explication of such structure.

In a tripartite hypergraph, the network \emph{\textbf{G}} can be
briefly depicted by
\emph{\textbf{G}}=(\emph{\textbf{V}},\emph{\textbf{H}}), where
\emph{\textbf{V}} denotes the vertices and \emph{\textbf{H}}
represents the set of hyperedges.
\emph{\textbf{V}}=\emph{\textbf{U}} $\cup$ \emph{\textbf{R}} $\cup$
\emph{\textbf{T}} where \emph{\textbf{U}}, \emph{\textbf{R}} and
\emph{\textbf{T}} represent the set of users, resources and tags
respectively, and \emph{\textbf{H}} $\subseteq$ {\emph{\textbf{U}}
$\times$ \emph{\textbf{R}}  $\times$ \emph{\textbf{T}}} is usually
much smaller than the number of all the possible triangles.
Correspondingly, the \emph{Del.icio.us} dataset we collected has
15009 users, 2431190 resources and 325120 distinct tags, which
subsequently constitute 11739998 hyperedges.

\subsection{Model}

Consequently, we are mainly interested in the effect of tagging
behaviors and the role of tags in networks. Therefore, we fix the
distribution of user activities according to the empirical data.
Thus, the model can be described as following:
\begin{itemize}
\item At each time step, pick up a random user $u$ according to the given distribution of
user activities.
\item For $u$, s/he can either choose a resource with
probability $p$, or select an arbitrary tag with probability 1-$p$.
\item If $u$ is activated from the aspect of resource, s/he will
randomly select an existing resource in the system with probability
1-$p_{1}$ according to its popularity, or introduce a completely new
resource with probability $p_{1}$. And then s/he will annotate it
with a few tags. For simplicity, in this paper, we only consider
that $u$ will assign only one tag to the selected resource $r$.
Thus, $u$ could choose the tag from his/her own vocabulary with
probability $p_{2}$ according to how many times s/he has adopted it
, or from the resource vocabulary with probability $p_{3}$ according
to how many times it has been associated with the target resource,
or introduce a new tag with probability 1-($p_{2}$+$p_{3}$) if s/he
does not find a suitable or personalized tag to describe $r$.
\item If $u$ decides to find a relevant resource from a specific
topic, s/he will choose a random tag $t$ based on its popularity,
and then save one of the relevant resources according to how many
triangles they have appeared together with $t$.
\end{itemize}

\begin{figure}[ht]
\center\scalebox{0.35}[0.35]{\includegraphics{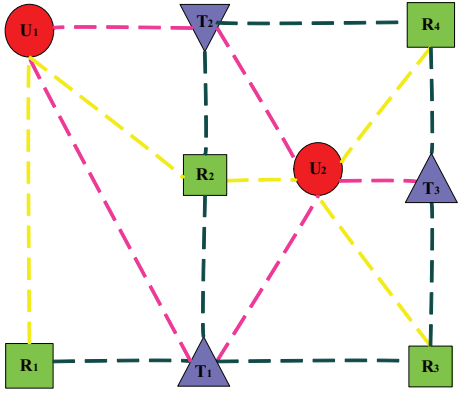}}
\caption{(Color online) A descriptive hypergraph consists of two
users, four resources and three tags. Take user $U_2$ and resource
$R_1$ for example, the measurements are denoted as: (i) $U_2$ has
participated six hyperedges, which means its hyperdegree is 6; (ii)
$U_2$ has directly connected to three resources and three tags,
which suggests it possibly has 3$\times$3=9 hyperedges in maximal.
Thus its clustering coefficient equals 6/9=0.667, where 6 is its
hyperdegree; (iii) the shortest path from $U_2$ to $R_1$ is
$U_2-T_1-R_1$, which indicates the distance between $U_2$ and $R_1$
is 2. }\label{measurements}
\end{figure}

In this model, a new hyperedge ($u$,$r$,$t$) is produced either from
the perspective of resources or tags at each time step. When one
tries to give a tag to a certain resource, s/he might choose a
previous tag s/he used before, or pick up one tag recommended by the
system. A new tag is added if no appropriate tags is available to
describe that resource. Thus a tag-growth mechanism is considered in
the present model. We then repeatedly run the model until enough
number of hyperedges is obtained. Moreover, we simply assume that
there is only one hyperedge emerges once the user is activated,
which is not the case in real networks. However, such simplified
assumption could help us examine the effects of different tagging
behaviors on the emergence of folksonomies. To evaluate our model,
we measure the following quantities (Fig. \ref{measurements} gives a
detailed description of these quantities):

 (i) \emph{hyperdegree distribution}: defined as the proportion that
 each hyperdegree occupies, where hyperdegree is defined as the number
 of hyperedges that a regular node participates in.

 (ii) \emph{clustering coefficients}: defined as the proportion of
 real number of hyperedges to all the possible number of hyperedges that a
 regular node could have.

 (iii) \emph{average distance}: defined as the average shortest path
 length between two random nodes in the whole network.

Since we are mainly interested in how the tagging behaviors
influence the emergence of folksonomies, we fix other parameters and
investigate the effect of $p$. In the following analysis, we set
$p_1$=0.3, $p_2$=$p_3$=0.45 as constants.

\subsection{Hyperdegree Distribution}
According to \cite{Ghoshal2009}, hyperdegree is defined as how many
triples a regular node takes part in. Thus we denote $p_{(k_u)}$,
$p_{(k_r)}$, $p_{(k_t)}$  as the fraction of users, resources and
tags, respectively. In terms of the model, $p_{(k_u)}$ is directly
derived from the empirical data. Therefore, we mainly focus on the
dynamics of $p_{(k_r)}$ and $p_{(k_t)}$. Firstly, we can write down
the rate equation for the distribution of resources
\cite{Albert2002} (In order to avoid confusion of the time symbol,
we use $l$ to represent the time in following descriptions):
\begin{equation}
\begin{array}{rcl}
p_{(k_r,l+1)}&=&p\{p_1p_{(k_r,l)}+(1-p_1)[(1-pr_{(k_r,l)})p_{(k_r,l)}
\\&& +(1-\delta_{k_r,1})pr_{(k_r-1,l)}p_{(k_r-1,l)}]\}
\\&& +(1-p)\{(1-\delta_{k_r,1})ptr_{(k_r-1,l)}p_{(k_r-1,l)}
\\&& +[1-ptr_{(k_r,l)}]p_{(k_r,l)}\}+\frac{1}{l}\delta_{k_r,1},
\end{array}
\end{equation}
where $p_{(k_r,l)}$ is denoted as the hyperdegree distribution of
resources at time $l$, $pr_{(k_r,l)}$ is the probability to pick up
an uncollected resource for $u$ with hyperdegree $k_r$ according its
popularity, $ptr_{(k_r,l)}$=$k_r/l$ is the probability to choose a
resource from a random tag $t$ at time $l$, and $\delta_{i,j}$ is
the Kronecker delta. The first brace shows the choice described in
the model, where the first term is the probability of adding a new
resource and the second term is the probability of selecting an
existing resource. The second brace depicts the evolutionary process
from the aspect of tags. However, it is not easy to identify the
distribution of each individual's absent resources, we
approximatively consider that distribution is direct proportion to
that of the system, that is,
\begin{equation}
  pr_{(k_r,l)} \approx \frac{k_r}{l}.
\end{equation}

\begin{figure*}[ht]
\centering
\includegraphics[width=0.325\textwidth]{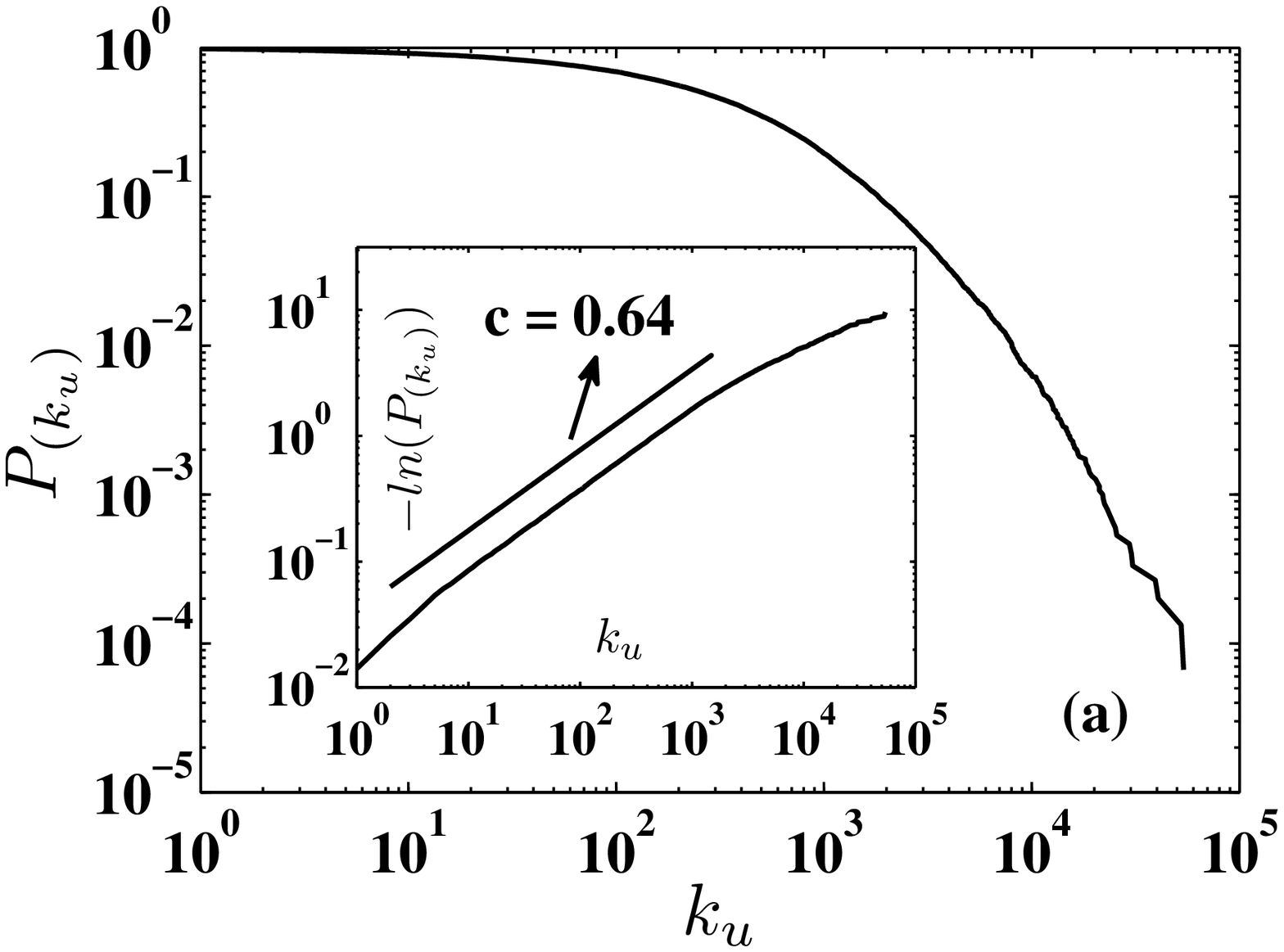}
\includegraphics[width=0.325\textwidth]{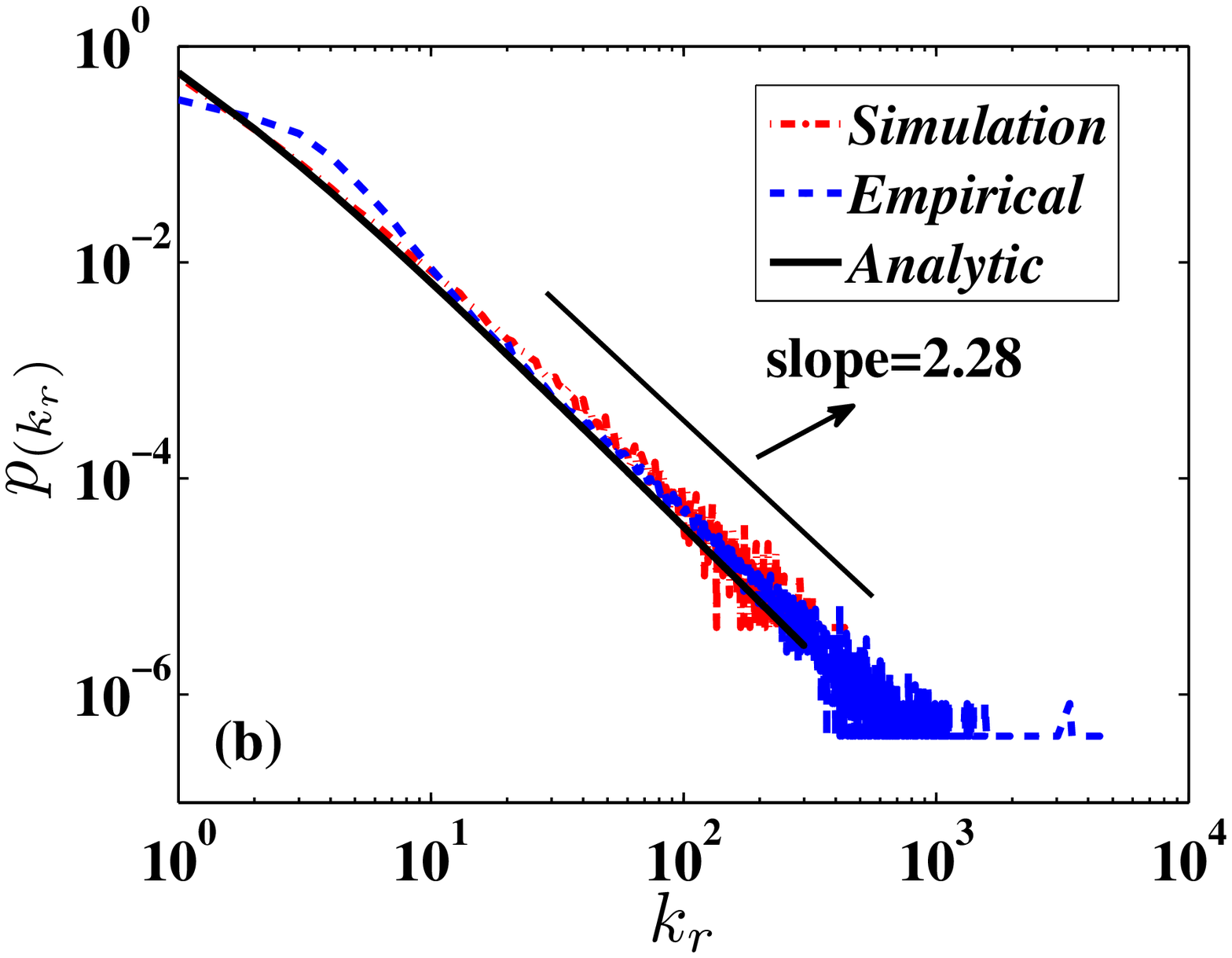}
\includegraphics[width=0.325\textwidth]{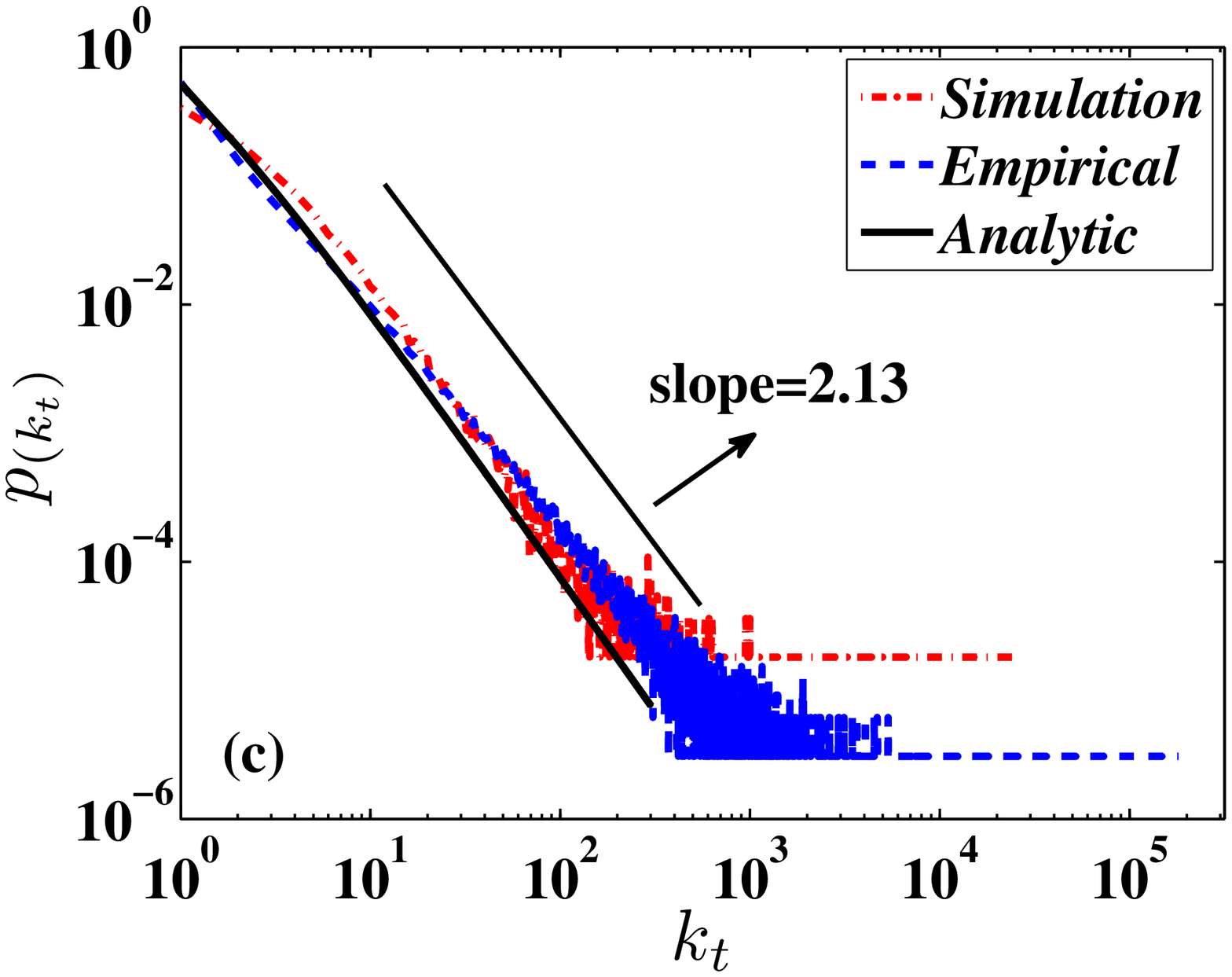}
\caption{(Color online) The hyperdegree distributions of three type
of nodes. (a) the empirical cumulative hyperdegree distribution of
users which follows a stretched exponential distribution $P{(k_u)}
\propto \exp^{-(k_u/k_0)^c}$, where $k_0$ is a constant. The inset
gives the fitting result of the exponent $c$=0.64 according to the
method used in \cite{ZhangPP2006}; (b) the empirical, simulation and
analytical results of resource hyperdegree distribution, following
power-low $p_{(k_r)} \propto k_r^{-\phi}$ and $\phi$=2.28; (c) the
empirical, simulation and analytical results of tag hyperdegree
distribution, following power-low $p_{(k_t)} \propto k_t^{-\varphi}$
and $\varphi$=2.13. The simulation and analytical results of (b) and
(c) are obtained when $p$=0.8.
   }\label{degree}
\end{figure*}

Integrate Eq. (1) and Eq. (2), as well as the stationary condition
$p_{(k_r)}=\mathop {\lim }\limits_{\scriptstyle l \to
\infty}\sum(p_{(k_r,l)})/l $, we have:
\begin{equation}
p_{(k_r)} \approx \left(\frac{k_r-1}{k_r+\frac{1}{1-pp_1}} \right)
p_{(k_r-1)},
\end{equation}
for $k_r$$>$1. When $k_r$=1, Eq. (1) can be simplified to:
\begin{equation}
p_{(k_r=1)}=\frac{1}{2-pp_1}.
\end{equation}

Combine Eq. (3) and Eq. (4), we can recursively obtain the final
solution:

\begin{equation}
p_{(k_r)}\approx
a_1\frac{\Gamma(k_r)\Gamma{(1+a_1)}}{\Gamma(k_r+1+a_1)},
\end{equation}
where $a_1$=$\frac{1}{1-pp_1}$ and $\Gamma$ is the Gamma function.

Analogously, we can also write down the tag hyperdegree distribution
in the form of rate equation:
\begin{equation}
\begin{array}{rcl}
p_{(k_t,l+1)}&=&[(1-p)+p(p_2+p_3)]
\\&&[pr_{(k_t-1,l)}p_{(k_t-1,l)}(1-\delta_{k_t,1})
\\&&+(1-pr_{(k_t,l)})p_{(k_t,l)}]
\\&&+p(1-p_2-p_3)p_{(k_t,l)}+\frac{1}{l}\delta_{k_t,1},
\end{array}
\end{equation}
where $pr_{(k_t,l)}$ is the probability of picking up a random tag
with hyperdegree $k_t$ at time $l$. According to the present model,
there are four mechanisms that drive the growth of tags: (i) user
$u$ selects one tag from his/her own vocabulary with probability
$p_2$; (ii) $u$ chooses one word from the set of tags associated
with the the target resource with probability $p_3$; (iii) a new tag
is introduced with probability 1-$p_2$-$p_3$; (iv) $u$ selects an
interesting tag $t$ from all the possible candidates and saves a
resource that is relevant with $t$. Eq. (6) exactly expresses the
integrated effect on tag evolution of those mechanisms.

We take the similar assumption of Eq. (2) that the individual's tag
hyperdegree distribution is direct proportion to that of the system:

\begin{equation}
  pr_{(k_t,l)} \approx \frac{k_t}{l}.
\end{equation}

We then follow the same processes of Eq. (3) and Eq. (4), the
solution will read:

\begin{equation}
p_{(k_t)}\approx
a_2\frac{\Gamma(k_t)\Gamma{(1+a_2)}}{\Gamma(k_t+1+a_2)},
\end{equation}
where $a_2$=$\frac{1}{1-p(1-p_2-p_3)}$.

Fig. \ref{degree} shows the simulation, analytical and empirical
results of hyperdegree distributions in both the real and modeled
networks. Fig. \ref{degree}(a) shows the empirical data of users'
cumulative hyperdegree distribution, which follows a stretched
exponential distribution \cite{Laherrere1998, Shang2010}. Fig.
\ref{degree}(b) and Fig. \ref{degree}(c) show good agreements among
empirical observation and analytical result, while the inconsistent
in Fig. \ref{degree}(b) might be caused by our assumption that
results in a comparatively large number resources with small
hyperdegrees. Note that $p=0.8$ indicates that most actions of the
tagging are from the resource aspect in folksonomies.

\begin{figure}[ht]
\centering
\includegraphics[width=0.4\textwidth]{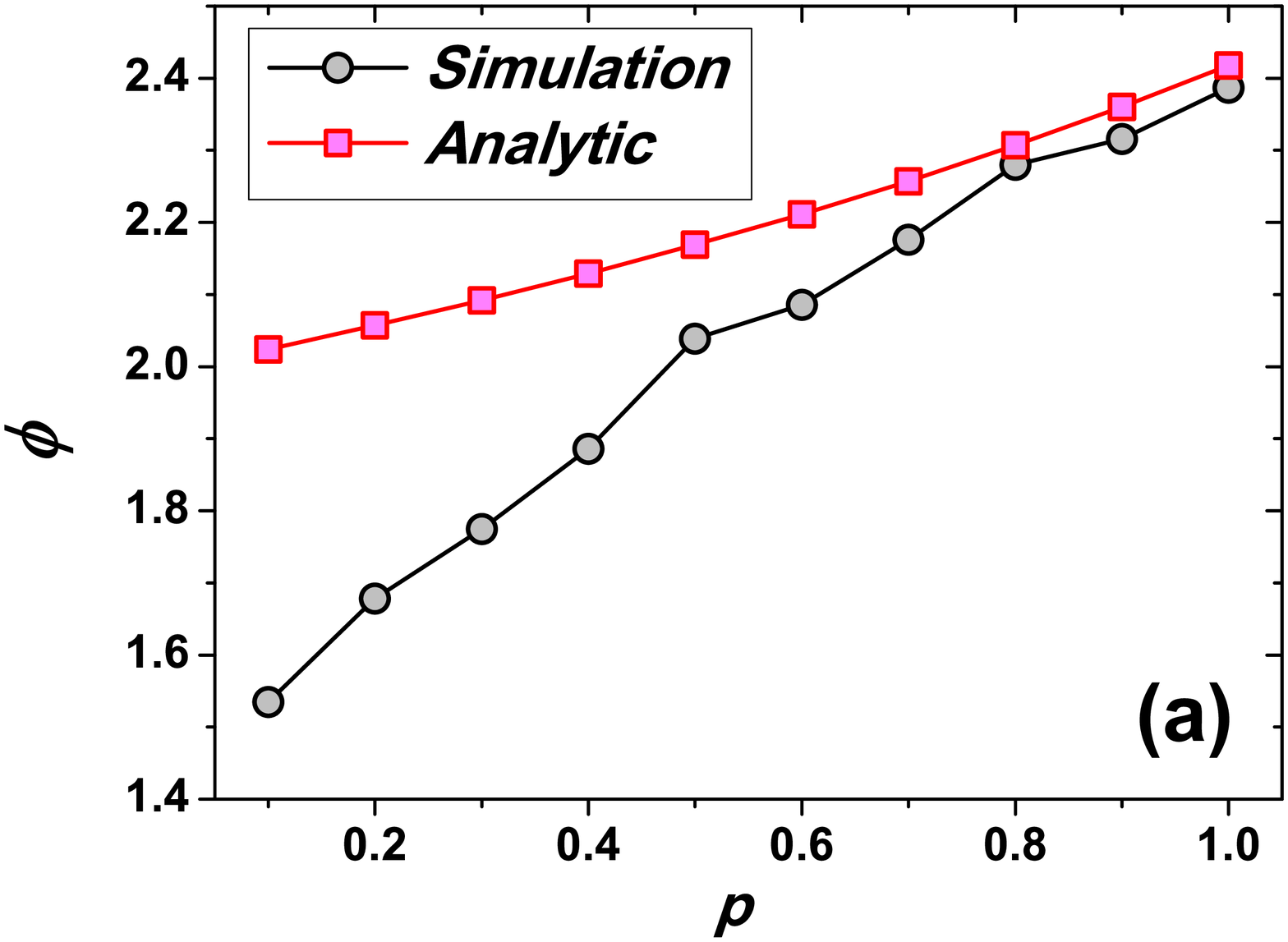}
\includegraphics[width=0.4\textwidth]{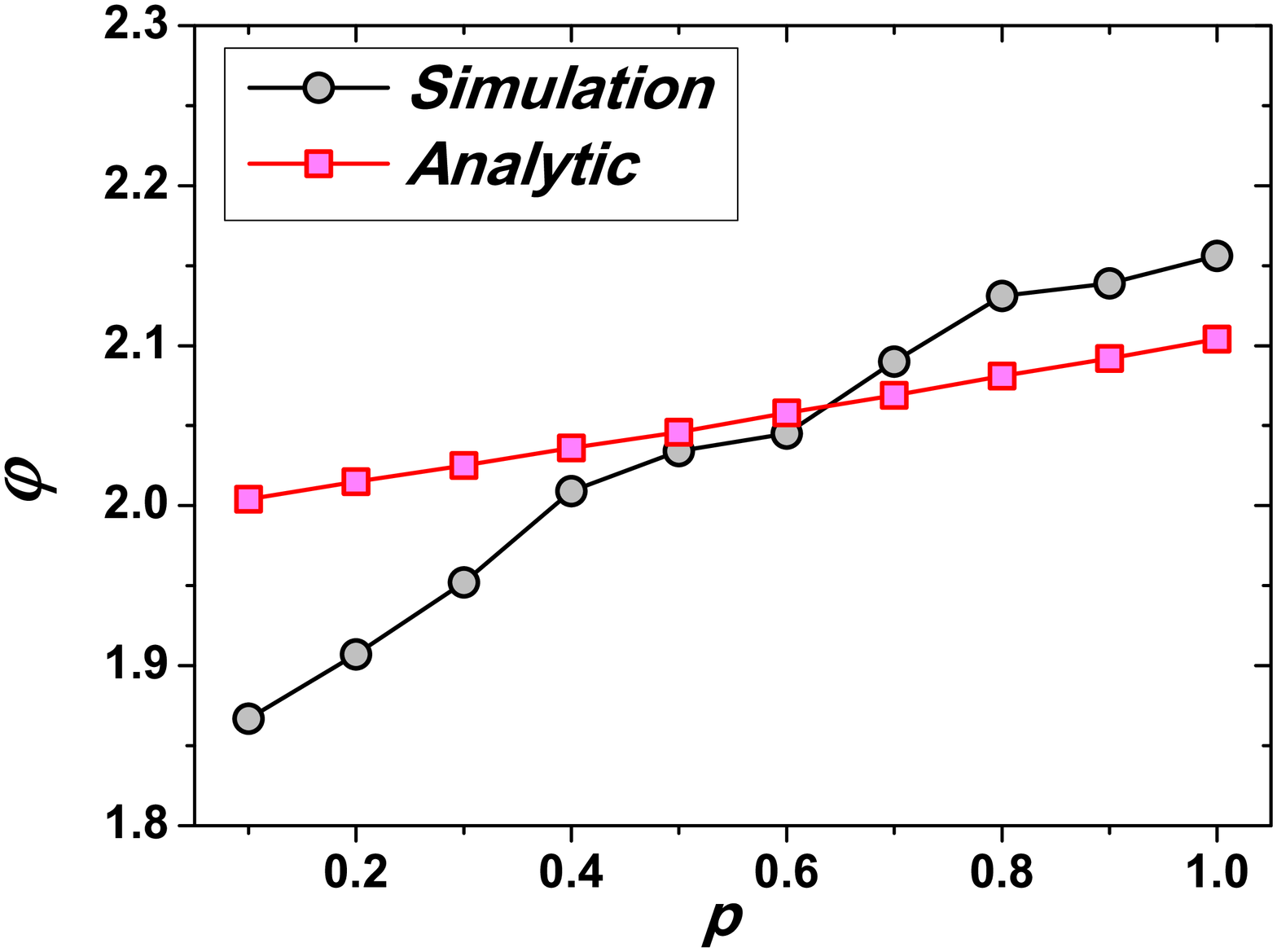}
\caption{(Color online) The slopes of hyperdegree distribution
change according to different value of $p$ for analytical and
simulation results. (a) the variation of $\phi$. (b) the variation
of $\varphi$. Both the two distributions show scale-free property
under disparate values of $p$, that is, $p(k) \propto k^{-\alpha}$,
where $\alpha$ refers to $\phi$ and $\varphi$ in (a) and (b),
respectively. $\phi$ and $\varphi$ are measured by Least Squares
Method (LSM).}\label{slope}
\end{figure}

In addition, we measure the effect on hyperdegree distribution with
different values of $p$. In Fig. \ref{slope}(a), the resource
hyperdegree distribution is in good agreement only when $p$
increases over 0.7. Whereas, the slope of tag hyperdegree
distribution does not change much with virous value of $p$. This
might be caused by two reasons: (i) the evolution of folksonomy is
driven primarily by assigning tags to the target resource, which is
consistent with large value of $p$; (ii) when $p$ is small, the
fat-tail of resources with small degree will remarkably affects the
fitting result.

\subsection{Clustering Coefficients}

\begin{figure*}[ht]
\centering
\includegraphics[width=0.325\textwidth]{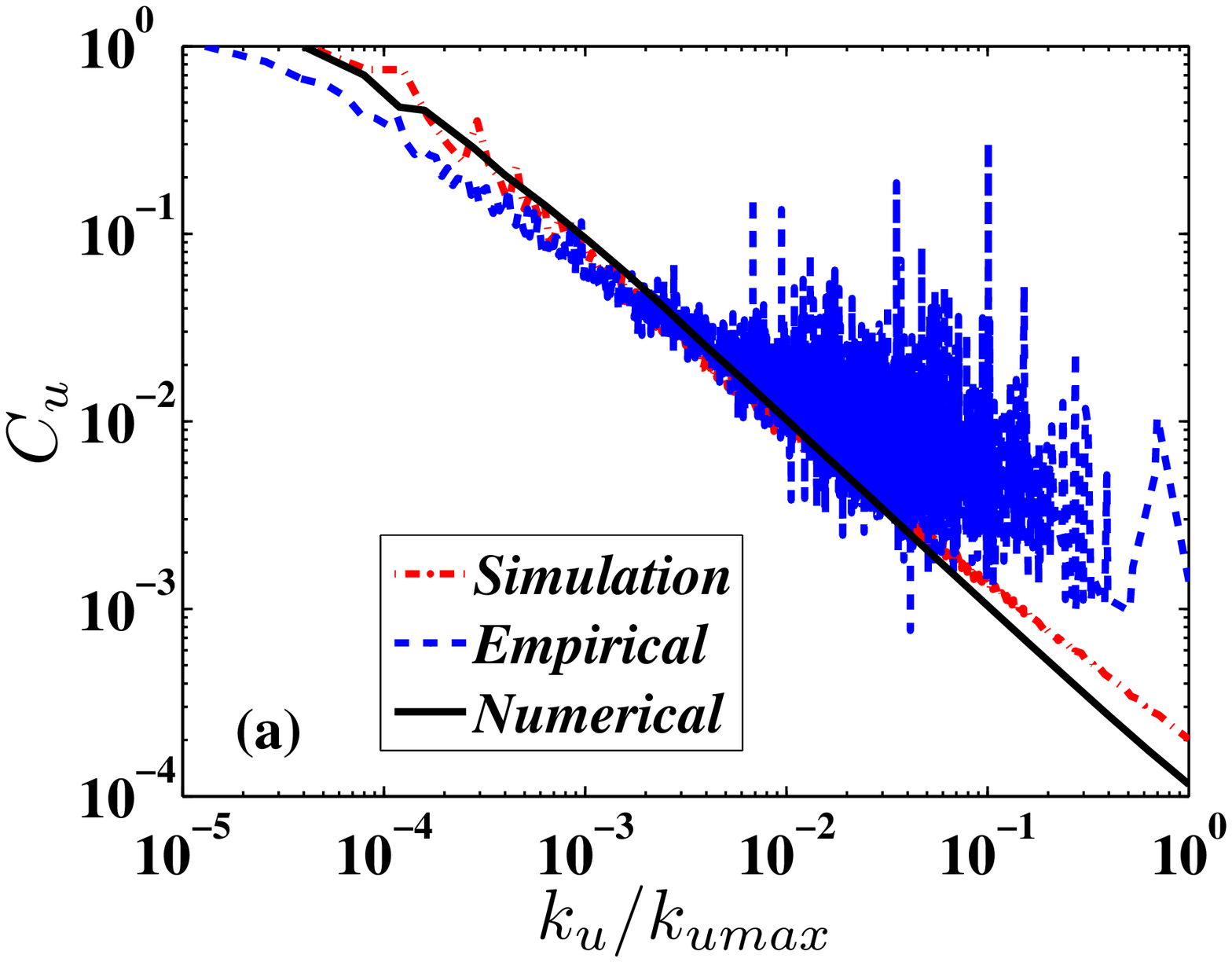}
\includegraphics[width=0.325\textwidth]{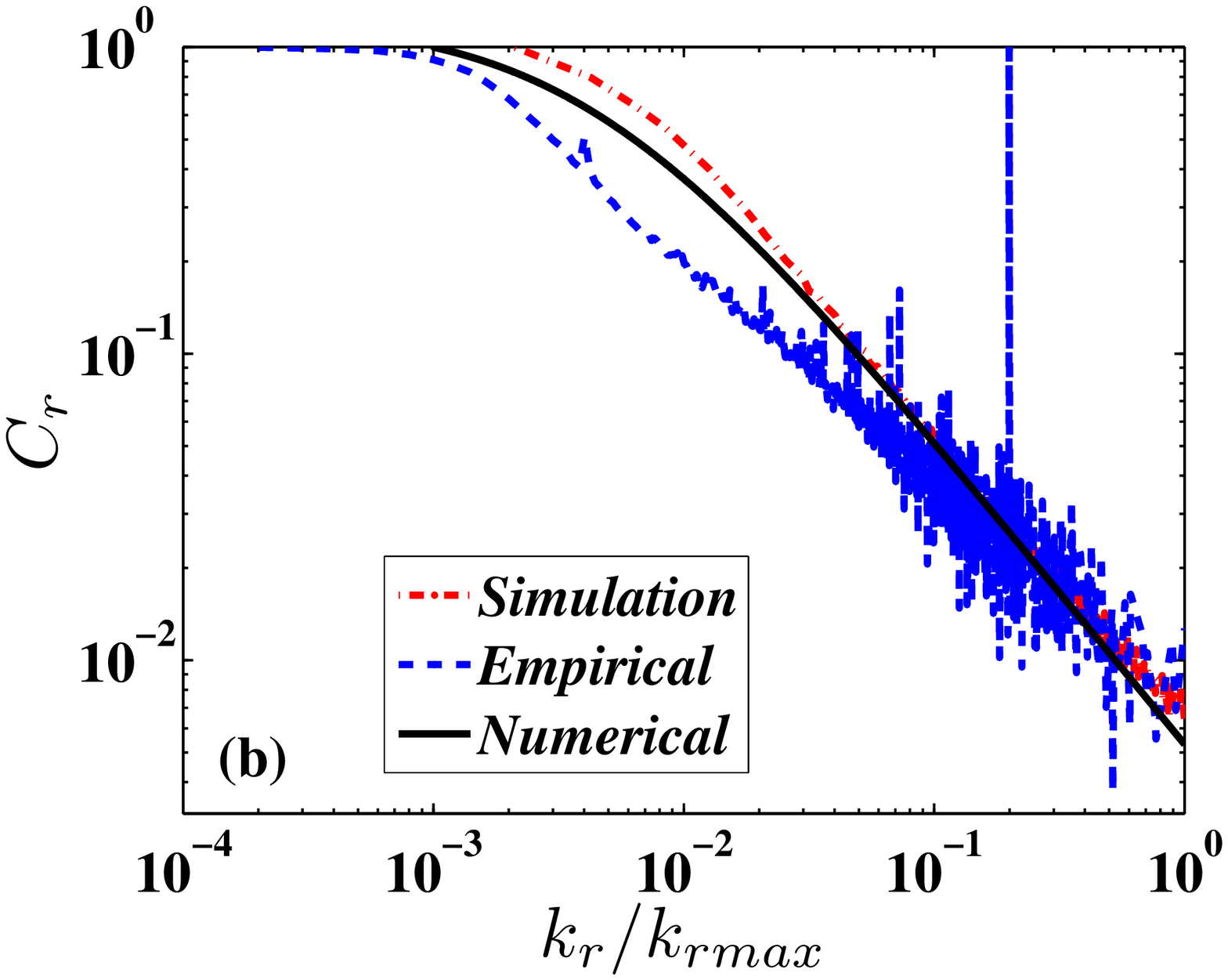}
\includegraphics[width=0.325\textwidth]{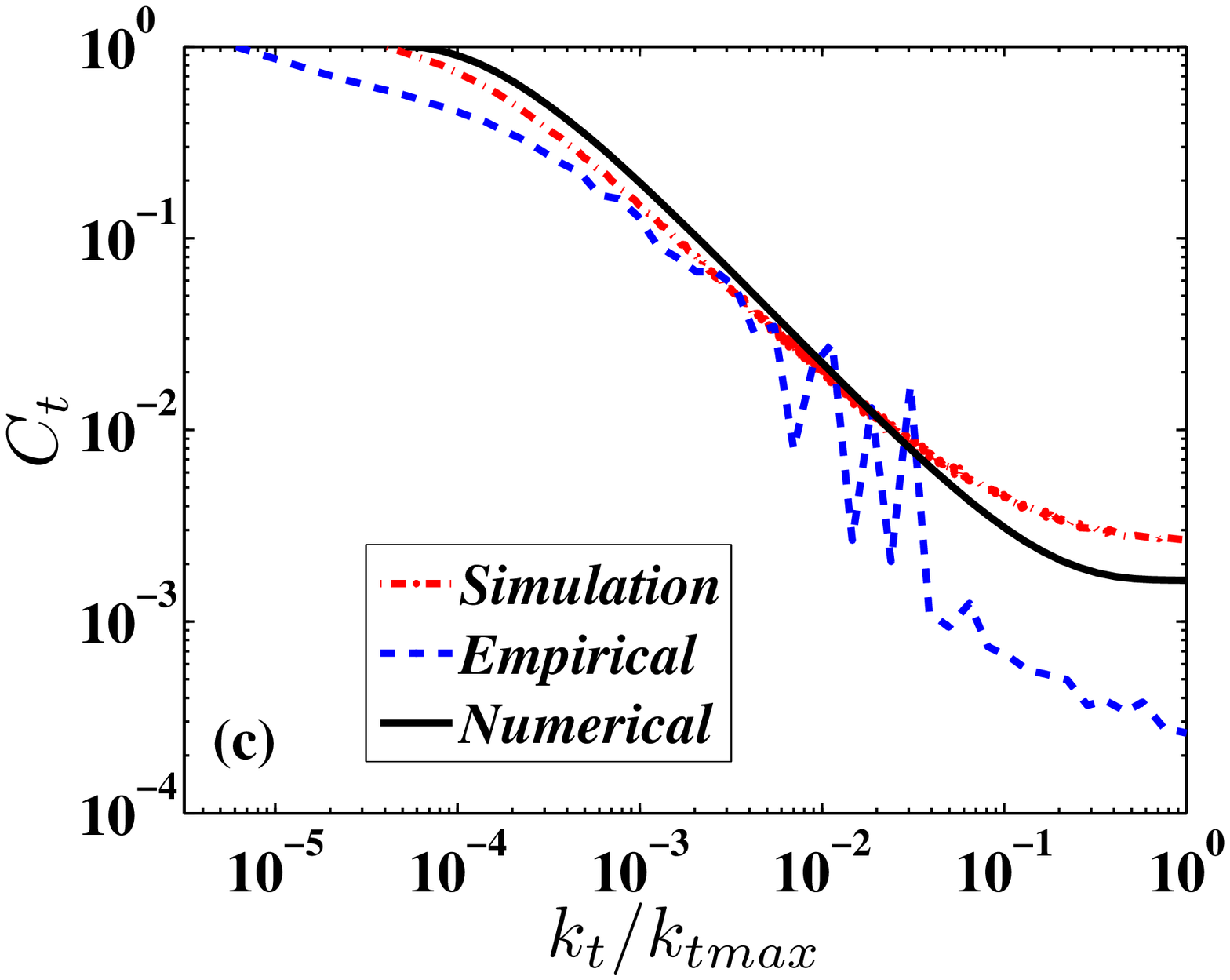}
\caption{(Color online) The clustering coefficients versus collapsed
hyperdegrees. (a) the user clustering coefficient versus collapsed
user hyperdegree; (b) the resource clustering coefficient versus
collapsed resource hyperdegree; (c) the tag clustering coefficient
versus collapsed tag hyperdegree. In (c), the empirical data is
shown in log bin in order to alleviate the fluctuation resulting
from insufficient data, 
interfering the exhibition of its statistical property. All the
three plots are obtained with $p$=0.8.} \label{cluster}
\end{figure*}

Clustering in a network measures the likelihood that two neighbors
of a given node are inclined to connect to each other. Watts and
Strogatz \cite{Watts1998} have introduced the \emph{clustering
coefficient} to measure the amount of clustering for a given node in
normal unipartite networks. However, this definition is not fully
compatible with the hypergraph case, since a regular node connects
two other different types of nodes. Thus, we adopt the definition of
user clustering coefficient given in Ref. \cite{Cattuto20072, f4}:
\begin{equation}
 C_u = \frac{k_u}{R_u \cdot T_u},\label{cluster_u}
\end{equation}
where $k_u$ is the hyperdegree of user $u$, $R_u$ is the number of
resources that $u$ collects and $T_u$ is the number of tags that $u$
possesses. The above definition measures the fraction of possible
pairs present in the neighborhood of $u$. A larger $C_u$ indicates
that $u$ has more similar topic of resources, which might also show
that $u$ has more concentrated on personalized or special topics.
Then the hyperdegree-based clustering coefficient is averaged over
all the nodes with the same hyperdegrees.

In order to compute $C_u$, we shall consider the evolutionary
dynamics of $T_u$, the number of tags used by the selected user, as
well as the dynamics of $T$, the current number of tags existing in
the system. We can write the differential functions:
\begin{equation}
 \left\{
 \begin{array}{rcl}
 \frac{d_{T_u}}{d_l}&=&\frac{k_u}{L}[pp_3(1-p_1)(1-\frac{T_u}{T})
 \\&&+p(1-p_2-p_3)(1-\frac{T_u}{T_0})
 \\&&+(1-p)(1-\frac{T_u}{T})],\\
 \frac{d_{T}}{d_l}&=&p(1-p_2-p_3)(1-\frac{T}{T_0}),
 \end{array}
 \right.
 \label{clustersolution_u}
\end{equation}
where $T_0$ is the total number of tags we initially set in the
model and $L$ is the total number of designed simulation steps.
Since we assume that only one tag is allowed to be assigned at each
time step, the hyperdegrees of users and resources are degenerated
to bipartite case. Therefore, we get $k_u$ = $R_u$. Thus, Eq.
(\ref{cluster_u}) can be rewritten as:

\begin{equation}
 C_u =  \frac{1}{T_u}. \label{clustersimplified_u}
\end{equation}

Unfortunately, It is not easy to get the explicit expression of Eq.
(\ref{clustersolution_u}). Instead, we find the numerical solution
by combining Eq. (\ref{clustersolution_u}) and Eq.
(\ref{clustersimplified_u}). Fig. \ref{cluster}(a) shows the good
consistency among the empirical, simulation and numerical results.

Analogously, we can also write the dynamics of $C_r$:
\begin{equation}
  \left\{
   \begin{array}{l}
   \frac{d_{T_r}}{d_l}=\frac{k_{r}}{l}p[p_2(1-p_1)(1-\frac{T_r}{T})+(1-p_2-p_3)(1-\frac{T_r}{T_0})],
   \\\frac{d_{T}}{d_l}=p(1-p_2-p_3)(1-\frac{T}{T_0}),
   \\\frac{d_{k_{r}}}{d_l}=\frac{k_{r}}{l},
   \\ C_r = \frac{k_r}{U_r \cdot T_r}= \frac{1}{T_r},
   \\
   \end{array}
  \right.
  \label{clustersolution_r}
\end{equation}
where $k_r$ is resource hyperdegree, $T_r$ is the number of tags
attached to resource $r$, and $U_r$ is the number of users who have
collect $r$. Fig. \ref{cluster}(b) shows the numerical solution for
Eq. (\ref{clustersolution_r}), as well as the empirical and
simulation results.

And the dynamics of $C_t$ is as following:
\begin{equation}
  \left\{
   \begin{array}{l}
   \frac{d_{U_t}}{d_l}=\frac{k_{t}}{l}[(1-p)(1-\frac{U_t}{U})+p_3(1-p_1)(1-\frac{U_t}{U})],
   \\\frac{d_{k_{t}}}{d_l}=\frac{k_{t}}{l}, \\
   \frac{d_{R_t}}{d_l}=\frac{k_{t}}{l}p[p_2(1-p_1)(1-\frac{R_t}{R})+p_1(p_2+p_3)],
   \\\frac{d_{R}}{d_l}=pp_1,
   \\C_t = \frac{k_t}{U_t \cdot R_t},
    \\
   \end{array}
  \right.
  \label{clustersolution_t}
\end{equation}
where $k_t$ is tag hyperdegree, $U$ is the number of users which is
fixed in the model, $U_t$ is the number of users who have used tag
$t$, $R$ is the number of resources existing in the system, and
$R_t$ is the number of resources labeled with $t$. Fig.
\ref{cluster}(c) shows the numerical solution for Eq.
(\ref{clustersolution_t}), as well as the empirical and simulation
results. All the three plots in Fig. \ref{cluster} show negative
correlations between clustering coefficient and hyperdegree on both
the real-world and modeled networks. It might indicate the
hierarchical structure of tripartite hypergraphs \cite{Ravasz2003},
and suggest that users with larger hyperdegrees have more diverse
interests, and vice verse.

\subsection{Average Distance}

\begin{figure*}[ht]
\centering
\includegraphics[width=0.4\textwidth]{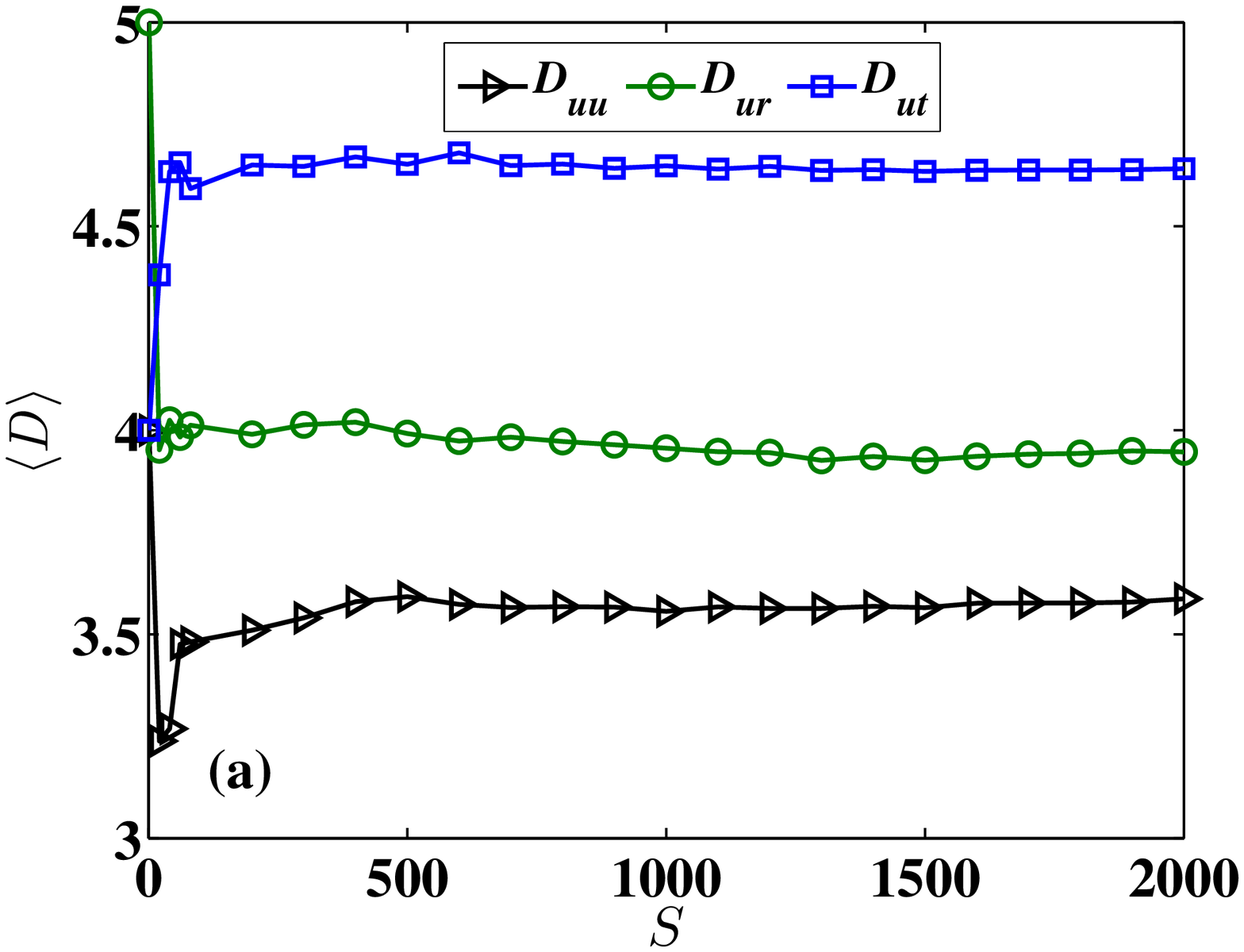}
\includegraphics[width=0.4\textwidth]{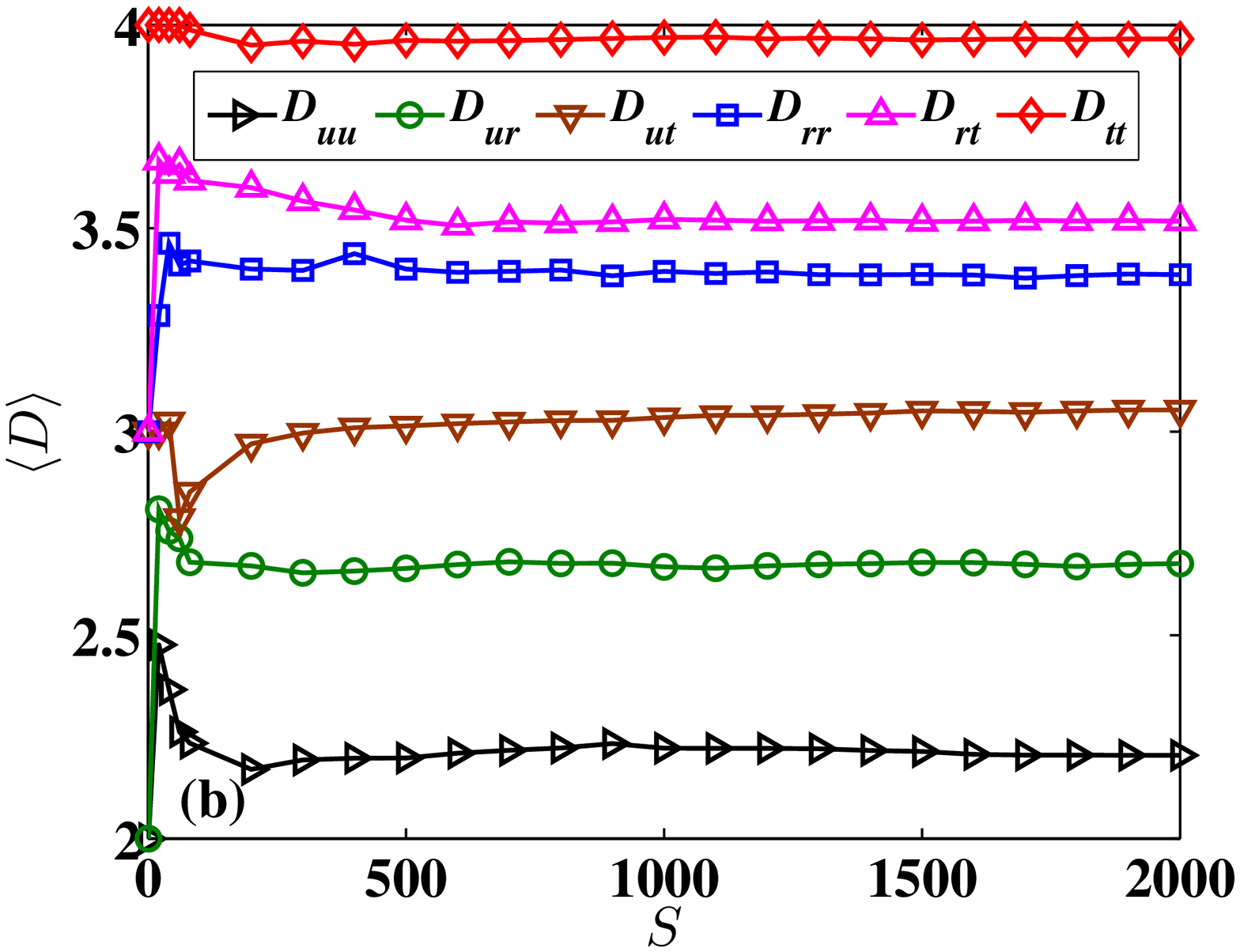}
\includegraphics[width=0.4\textwidth]{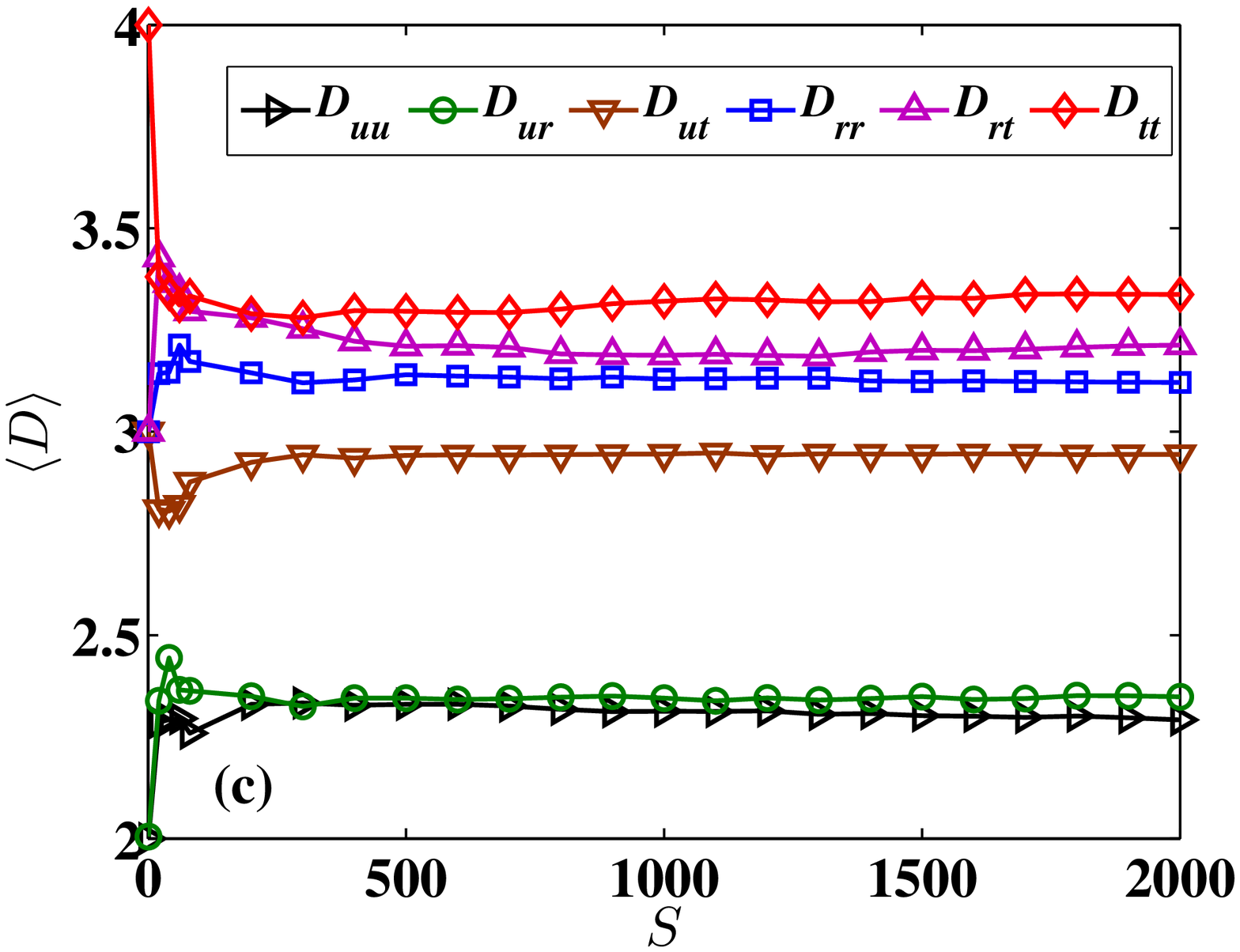}
\includegraphics[width=0.4\textwidth]{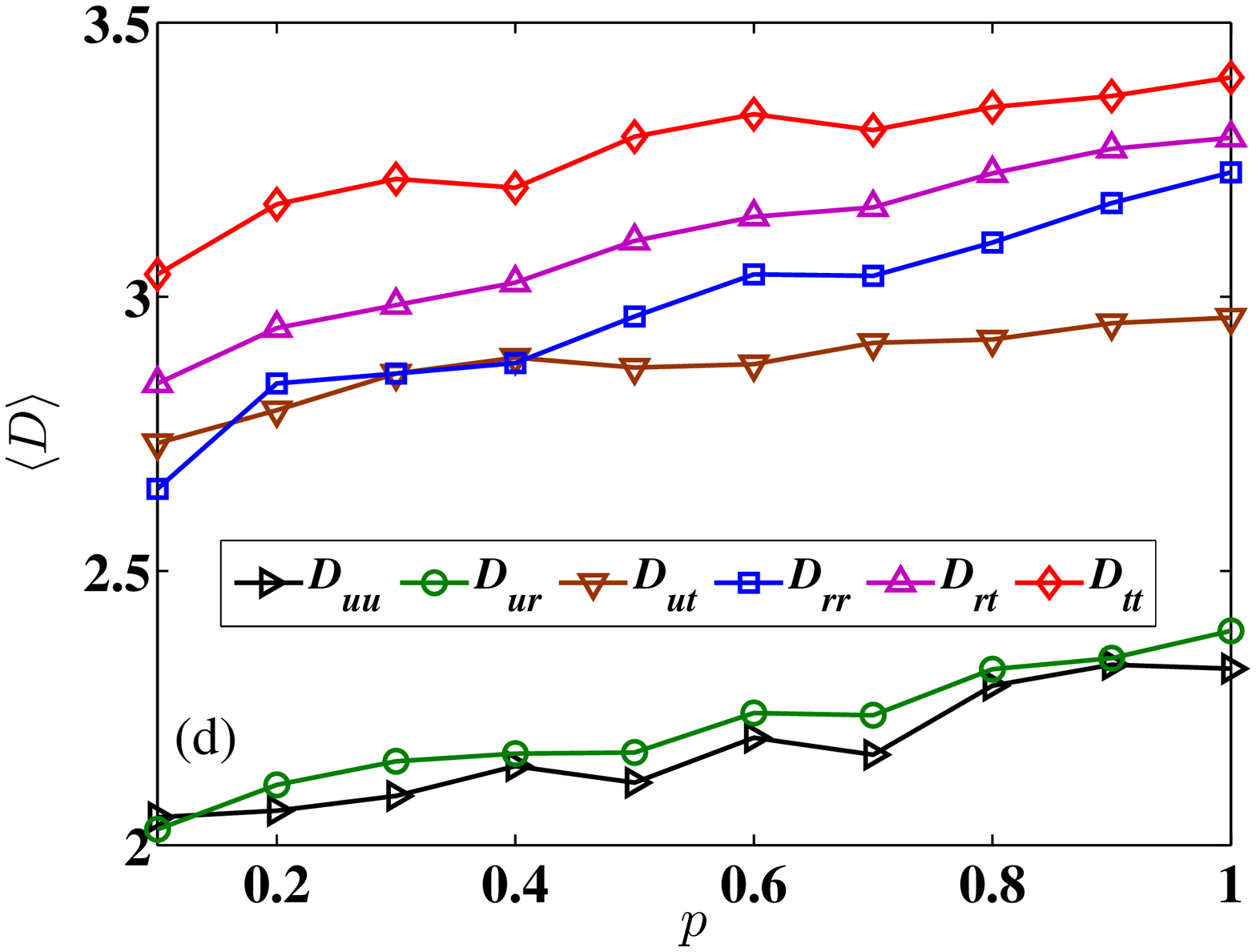}
\caption{(Color online) The average distances of bipartite and
tripartite networks. Since the data is huge, we calculate $\langle
D\rangle$ by sampling randomly pairs of nodes until stationary value
is obtained. (a) the average distances of user-user ($D_{uu}$),
user-resource ($D_{ur}$) and resource-resource ($D_{rr}$) versus the
number of samplings in the bipartite network, ignoring the tag
information of \emph{Del.icio.us}; (b) the average distances of
user-user, user-resource, user-tag ($D_{ut}$), resource-resource,
resource-tag ($D_{rt}$) and tag-tag ($D_{tt}$) versus the number of
samplings in the tripartite hypergraph of \emph{Del.icio.us}; (c)
the average distances of user-user, user-resource, user-tag,
resource-resource, resource-tag and tag-tag versus the number of
samplings in the tripartite hypergraph produced by the present
model. All the curves converge fast with just a small number of
samplings, which indicates a small-world property in both bipartite
and tripartite networks; (d) the stationary average distances change
according to different values of $p$ in the modeled
network.}\label{distance}
\end{figure*}

Another important quantity is the distance, $D$, between a random
pair of nodes in a network. Hence, the average distance, $\langle
D\rangle$, measures the efficiency of retrieving a target node in a
network. Take a friendship network for example, $\langle D\rangle$
is given by counting the average shortest path length between a
random user and another arbitrary user. Therefore, $\langle
D\rangle$ assesses how easily yet effectively for a user to make
acquaintance with others in a given friendship network.

However, in the case of tripartite hypergraph, there are three
different regular nodes. Therefore, the shortest path length can be
defined as the minimal number of hyperedges that must be traversed
to go from vertex to vertex. Fig. \ref{distance} shows the $\langle
D\rangle$ between any two types of vertices. Fig. \ref{distance}(a)
and Fig. \ref{distance}(b) show the average distances of the
bipartite network and hypergraph structure of \emph{Del.icio.us},
respectively. We can see that: (i) tags can significantly shorten
$\langle D\rangle$ for any pair of nodes in comparison with the
bipartite case. For example, $\langle D\rangle$ of user-user pair is
enhanced from 3.587 to 2.205, $\langle D\rangle$ of user-resource
pair is improved from 3.947 to 2.676, and the value of $\langle
D\rangle$ of resource-resource pair is shortened from 4.641 to
3.386. These considerable improvements might indicate that tags play
an important role in \emph{Information Retrieval}; (ii) in Fig.
\ref{distance}(b), the magnitude strictly follows the order:
$D_u<D_r<D_t$ in both general and special cases. For example, we
have: $D_{uu}<D_{ur}<D_{ut}$ for users, $D_{ur}<D_{rr}<D_{rt}$ for
resources, and $D_{ut}<D_{rt}<D_{tt}$ for tags. The similar patten
of those orders might imply that \emph{Del.icio.us} is a
user-centric system so that we can more easily find any information
through users than others. Besides, the main purpose of tagging is
to more efficiently and effectively manage resources, which keeps
coherence of comparatively large value of $p$ in previous sections.
Fig. \ref{distance}(c) reproduces such exciting phenomenon with
$p$=0.8 in the model. Furthermore, we study the effect of different
values of $p$ on the distances. In Fig. \ref{distance}(d), it is
shown that the order does almost keep steady whatever the value $p$
changes to. Additionally, Fig. \ref{distance}(d) also indicates that
all the distances decreases monotonously with the lessening of $p$,
which might suggest that the more often we use tags, the more
effective we can find target information.

\section{Conclusion and Discussion}
In this paper, we have proposed an evolutionary hypergraph model to
study the dynamical properties of social tagged networks, so-called
folksonomies. The present model assumes that there are two typical
tagging behaviors based on preferential attachment mechanism: (i)
assigning tags to users' favorite resources; (ii) saving resources
that are relevant to interesting tags. The resulting tripartite
hypergraph shows good agreement with a real-world network,
\emph{Del.icio.us}, in following aspects: (i) the power-law
hyperdegree distributions are generated for resources and tags,
which indicates the heterogeneous topology; (ii) the decay of
average clustering coefficients with the increase of hyperdegree,
which may indicate hierarchical structure of tripartite hypergraphs
; (iii) the average distances between vertices of hypergraph are
comparatively smaller than those in corresponding bipartite networks
without tags; (iv) the relatively small average distance indicates a
small-world property, which facilitates the \emph{serendipitous
discovery} of interesting contents and congenial companions; (v) all
the above properties are found relatively high consistency with a
comparatively large value of $p$=0.8, which suggests that the
majority of actions is motivated by the first tagging behavior.
Consequently, this model quantitatively reveals the accessorial yet
significant role that tags play in folksonomies.

However, despite the good agreements in reproducing several features
with real data, it is not easy to fully uncover the mechanisms
dominating the emergence of folksonomy. This paper only provides a
start point for understanding the underlying motivations in
facilitating a variety of intricate properties in such new
paradigms. The present model considers that only one hyperedge is
allowed to come forth at each time step, which is moderately not the
case in real systems.  The tag co-occurrence \cite{Cattuto20071,
Cattuto20072} and social cognitive imitation mechanisms
\cite{Dellschaft2009} can be taken into account to improve the
proposed model.

\section*{ACKNOWLEDEMENTS} We acknowledge Dong Wei
for providing us the data set, Jian-Guo Liu, Linyuan L\"{u}, Chi-Ho
Yeung and Tao Zhou for helpful discussions and suggestions. This
work is partially supported by the Swiss National Science Foundation
(Project 200020-121848). ZKZ acknowledges the National Natural
Science Foundation of China under the grant nos. 60973069 and
90924011. CL and ZKZ acknowledge the Scholarship Program supported
by China Scholarship Council (CSC Program).

\end{document}